\documentstyle[prb,aps,epsfig]{revtex}
\begin{document}
\twocolumn
\title{Static NLO susceptibilities: \\testing approximation schemes against
exact results}

\author{Luca Del Freo, Francesca Terenziani, and 
Anna Painelli\thanks{Corresponding author:
Dipartimento di Chimica GIAF,
Universit\`a di Parma, Viale delle Scienze 17/A, I--43100, Parma, Italy.
Tel. +39--0521--905461. Fax. +39--0521--905556. 
E-mail: anna.painelli@unipr.it}} 

\address{Dipartimento di Chimica Generale ed Inorganica, Chimica Analitica
  e Chimica Fisica \\ Universit\`{a} di Parma, I--43100 Parma, Italy}

\date{\today}
\maketitle

\begin{abstract}
The reliability of the approximations commonly adopted in the calculation of 
static optical (hyper)polarizabilities is tested against  exact results 
obtained for an interesting toy-model. 
The model accounts for the principal features of 
typical nonlinear organic materials with mobile electrons strongly coupled
to molecular vibrations.
The approximations introduced in sum over states and finite field
schemes are analyzed in detail. Both the Born-Oppenheimer and the clamped 
nucleus approximations turn out to be safe for molecules, whereas
for donor-acceptor charge transfer complexes deviations from 
adiabaticity are expected. In the
regime of low vibrational frequency, static susceptibilities are strongly 
dominated by the successive derivatives of the potential energy and
large vibrational contributions to hyperpolarizabilities are found.
In this regime anharmonic corrections to hyperpolarizabilities are
very large, and the harmonic approximation, exact for the linear 
polarizability, turns out  totally inadequate for nonlinear responses. 
With increasing phonon frequency
the role of vibrations  smoothly decreases, until, in the antiadiabatic 
(infinite vibrational frequency) regime, vibrations do not contribute 
anymore to static susceptibilities, and the purely electronic 
responses are regained.
 
\end{abstract}

\narrowtext

 \section{Introduction}

Photonic and optoelectronic applications heavily rely on materials with
high nonlinear optical responses, and the development of new and more
efficient materials for nonlinear optics (NLO) is a key issue 
in present days research.
In this context, organic materials play an important role in view of
their large and fast NLO responses and of the tunability of
their properties via chemical synthesis.\cite{marder,bredas,special,rumi}
In order to trace guide-lines to organic synthesis, much  effort
is devoted to understand the origin of nonlinearity in these systems
and to relate NLO responses to electronic structure
and molecular geometry.\cite{special,rumi,bozio,zerbi,balakina}
Mobile electrons are required for large nonlinearity and 
conjugated molecules and/or polymers are materials of choice for NLO
applications.\cite{bredas,special} 
Vibrations are strongly coupled to conjugated 
electrons:\cite{soos} the effect of electron-phonon (e-ph) coupling 
in NLO responses at optical frequencies is 
not fully understood yet\cite{berkovic,wang,chernyak,bishop}, but general 
agreement has emerged on the importance
of e-ph coupling in static NLO responses.\cite{kim,bishop98,painelli98}

Static susceptibilities are the opposite of the successive derivatives of the
ground state (gs) energy with respect to a static electric field ($F$):
\begin{equation}
{\mathcal{E}}={\mathcal{E}}_{0}-\mu F-\frac{1}{2}\alpha F^2-
	\frac{1}{6}\beta F^3-\frac{1}{24}\gamma F^4+ \dots
\label{mu}
\end{equation}
where ${\mathcal{E}}_{0}$ is the gs energy
at zero field, $\mu$ is the permanent dipole moment, 
$\alpha$ is the linear polarizability, and $\beta$ and $\gamma$ are the first and
second hyperpolarizabilities, respectively.
As it is always the case
for energy derivatives, two different methods can be adopted to evaluate 
susceptibilities.\cite{bishop98}
The first method, the so-called 
finite-field (FF) approach, basically relies on direct 
numerical derivation techniques. 
The relevant  system is the molecule in the presence
of an external static electric field.\cite{cohen} 
The corresponding Hamiltonian is diagonalized to calculate the
gs energy (or dipole moment) at different $F$ values, and 
(hyper)polarizabilities are calculated as numerical derivatives
of the energy (or dipole moment) with respect to the electric field.
The second approach treats $F$ as a perturbation 
on the molecular states, leading to
 the well-known sum over states (SOS) formulas.\cite{orr}
The main advantage of SOS over FF is that
only unperturbed ($F=0$) eigenstates are required.  The main 
disadvantage is  that all  eigenstates and  transition dipole moments
are required, whereas FF only requires the gs energy
(or dipole moment).
Of course the two approaches are perfectly equivalent
if exact energies and eigenstates,
including vibrational contributions, are inserted into equations.
But for molecules and, more generally, for molecular materials, 
exact vibronic eigenstates are hardly accessible, and several
approximations are introduced.
Approximated calculations based on FF or SOS approaches are clearly different
and it is important to investigate the merit and reliability of the various
approximation schemes.

The first approximation usually introduced
is the adiabatic (Born-Oppennheimer, BO) approximation.
This is an almost ubiquitous approximation in molecular and condensed matter
physics: accounting for the different masses of electrons and nuclei (and then
for the different time scales of their motions),  BO approximation separates
 electronic and vibrational degrees of freedom, so that each state 
is factorized into a product of an electronic and a vibrational wavefunction.
This approximation is fairly safe: deviations from
the adiabatic regime are expected only 
for degenerate or quasi-degenerate states, for narrow-band solids,\cite{ciuchi}
or near to phase transitions.\cite{borghi}

In SOS approaches  BO factorization
 is used to separate electronic and vibrational contributions to 
susceptibilities.
Electronic contributions are those where all summations
run on vibronic states in the  excited electronic manifolds,
 whereas vibrational 
contributions have at least one of the indices in the summation running over
excited vibrational states within the gs manifold.\cite{bishop91}
In the same spirit as BO, clamped nucleus (CN) approximation has been proposed
to calculate SOS expressions for susceptibilities 
in a two-step procedure.\cite{bishop90}
In the first step the electronic (hyper)polarizabilities are calculated
with nuclei clamped at their equilibrium positions.
The resulting electronic contributions to (hyper)polarizabilities only contain
sums over purely electronic states.
In the second step the nuclear motion on the gs potential energy surface (PES) 
is accounted for, and
vibrational contributions to susceptibilities are obtained.
The adiabatic approximation, 
as directly applied to SOS  or within the CN
formalism, makes the calculation of susceptibilities possible,
provided that electronic and vibrational eigenstates are known.
However, in view of the anharmonicity of the relevant PES, calculating
the vibrational states in all electronic manifolds (as required in BO-SOS 
formalism) or just in the gs manifold (as required in CN-SOS) is not easy.
Often SOS calculations of static (hyper)polarizabilities
of molecules and/or polymers are carried out in the harmonic 
approximation,\cite{kim,yaron,champagne94,bishopjcp} where the solution
 of the vibrational problem can be written in closed form.
Harmonic schemes work in the hypothesis that the 
electronic system responds linearly to the nuclear motion: harmonic
approximation is  bound to fail in systems where vibrations are coupled
to mobile electrons, i.e. to electrons characterized by large nonlinear
responses.\cite{delfreo,understanding}
In fact large anharmonic corrections to NLO responses have been recognized in 
recent {\it ab initio} calculations on conjugated molecules and 
polymers.\cite{champagne00,luis}

When applied to FF approaches, BO quite naturally leads to a 
two-step procedure. First
of all the adiabatic electronic Hamiltonian (i.e. the total 
Hamiltonian minus the nuclear kinetic energy, KE, term) is solved to get
the gs electronic energy, that represents the 
potential energy (PE) for the nuclear motion. 
Its derivatives on $F$ measure the PE contributions  to  susceptibilities.
The calculation of the PE contribution   is easy since  only 
the gs electronic energy is required. 
Moreover, if nuclei are allowed to relax in response to $F$, 
susceptibilities calculated as PE derivatives naturally include  vibrational 
contributions, and fully account  for anharmonicity. 
The second step accounts for KE contribution: this is
a more difficult step because the energy of the lowest vibrational state
has to be calculated as a function of $F$, a
 non-trivial task due to anharmonicity.
Moreover, in systems with large e-ph coupling, like conjugated polymers and/or 
oligomers, it can be difficult to get reliable calculations of $F$-dependent 
geometries, and numerical instabilities of the  energy derivatives 
 are sometimes observed.\cite{ingamells}

In this paper we critically review the different approximation
schemes for the calculation of static NLO responses, 
when applied to molecules of interest for NLO applications. In these molecules,
in fact, electron-phonon coupling has particularly large and
non-trivial effects, shedding doubts on the reliability of approximations that
work well in more conventional systems.
In particular we will focus on an interesting toy-model for mobile electrons
 coupled to vibrations: the Holstein $DA$ dimer.
The so-called $DA$ dimer is a two-state model originally proposed by
Mulliken to describe electron donor-acceptor charge 
transfer (CT) complexes in solution.\cite{mulliken}
 Later on it was applied by Oudar and Chemla\cite{oudar} to
describe push-pull chromophores where the donor and acceptor moieties 
are linked by a $\pi$-conjugated bridge to form a molecule. 
The extension of the model to include
Holstein coupling to vibrations was first suggested by 
Person et al.\cite{person} to describe
vibrational spectra of CT complexes, and was extensively discussed in the 80's
to describe  many  properties (ranging from vibrational spectra to structural 
instabilities) of CT salts.\cite{soos2,bibbia,pgprb} 
Holstein coupling in the $DA$ dimer has also been adopted to describe
vibrations and/or polar solvation effects in push-pull 
chromophores\cite{painelli98}: an extensive comparison with spectral 
data for several key molecules proved that this model  contains 
the main ingredients to understand 
nonlinearity in these systems.\cite{understanding,terencpl,jpc}
In spite of its richness, the model is very simple
and in the next Section we will describe its exact, non-adiabatic 
eigenstates, as well as its BO solution. 
Sections III and IV discuss the approximations usually introduced in 
SOS and FF approaches, respectively; Section V summarizes main results.

\section{The Holstein $DA$ dimer: non-adiabatic and adiabatic eigenstates} 

The $DA$ dimer describes two electronic states,
$|DA\rangle$ and $|D^+A^-\rangle$, linearly coupled to a harmonic phonon, 
as depicted in Fig.~1. 
The relevant Hamiltonian is ($\hbar =1$):
\begin{equation}
{\mathcal{H}} = 2z_0\hat{\rho}-\sqrt{2}t\hat{\sigma}_x+
\frac{1}{2}\left(\omega^2Q^2+P^2\right)
-\sqrt{2\epsilon_{sp}}\omega Q\hat{\rho}-\hat{\mu}F.
\label{hamiltonian}
\end{equation}
The first two terms in the above equation describe the electronic Hamiltonian, 
where $\hat{\rho} =(1-\hat{\sigma}_z)/2$ is the ionicity operator,
 and $\hat{\sigma}_x$, $\hat{\sigma}_z$
represent the Pauli spin operators.
The third and fourth terms describe a 
harmonic vibration of frequency $\omega$ ($Q$
and $P$ are the corresponding coordinate and momentum), coupled to the 
electronic system. The  strength of the coupling is 
 measured by $\epsilon_{sp}$, the relaxation 
energy of $|D^+A^-\rangle$ (see Fig.~1a). 
The last term accounts for a static electric field, $F$, interacting with
the dipole moment operator, defined,
by $\hat{\mu}=\mu_0\hat{\rho}$, with $\mu_0 = 
\langle D^+A^-|\hat{\mu}|D^+A^-\rangle$.\cite{mulliken} 
In the following, energies are expressed in $\sqrt{2} t$ units.
We underline that the single-mode Hamiltonian in the above equation captures
the essential physics of e-ph coupling, at least if one is not interested in 
the detailed description of vibrational spectra.\cite{painelli98,pgprb,terencpl}
The extension to the multimode case, possible in principle, is very demanding
if the non-adiabatic solution is required.

The above Hamiltonian can be numerically diagonalized on the basis of the
direct product of the two electronic
states $|DA\rangle$ and $|D^+A^-\rangle$ and of the reference 
vibrational states (i.e. the eigenstates of the
harmonic oscillator in the third term of Eq.~(\ref{hamiltonian})).\cite{delfreo}
The  basis is truncated by fixing a maximum number
of phonon states, $M$; the corresponding $2M\times 2M$ matrix  can be 
diagonalized  up to  fairly large $M$ values, yielding numerically exact 
non-adiabatic eigenstates. The minimum $M$  required
to get convergence depends on the model parameters and on
the  properties of interest.
All results presented in the following have been obtained with $M=50$.  

The exact non-adiabatic eigenstates at $F=0$ are inserted into standard SOS 
expressions for static susceptibilities\cite{delfreo} to get the exact NLO 
responses  reported as continuous lines 
in Fig.~2 for $\epsilon_{sp} =1$ and a few $\omega$ values. By the way, 
exactly the same results are obtained, within  FF approach,
 from the  successive derivatives of the exact 
non-adiabatic gs energy vs the applied field. 
In the same figure, dot-dashed lines correspond to the bare electronic 
susceptibilities, $\chi_0^{(n)}$, i.e. to the response of the two-state model
with no e-ph coupling. The deviations of the continuous lines from
the dot-dashed lines measure the vibrational contribution to static responses.
Dashed lines show the  PE contribution to susceptibilities.
Analytical expressions for these curves  have already
been reported in Ref.~\cite{painelli98}: PE-susceptibilities are independent
of $\omega$. Fig.~2 clearly shows the evolution of static susceptibilities with
phonon frequency: 
in the low-$\omega$ limit, KE contributions vanish, and the exact curves
tend to the limiting PE results. Vibrational contributions to static 
susceptibilities are very large in this limit, and increase with the order
of nonlinearity. 
With increasing $\omega$, the vibrational contributions to static NLO
responses decrease: in the antiadiabatic limit ($\omega \rightarrow
\infty$) the vibrational contributions to static hyperpolarizabilities
vanishes, and the  exact curves tend to the bare electronic responses.

For real  molecules or complexes, the large number of electronic states 
makes non-adiabatic calculations very demanding,
and BO approximation is usually invoked.  
In  BO approximation the effective electronic Hamiltonian, 
${\mathcal{H}}_{el}={\mathcal{H}}-1/2P^2$, is defined by
subtracting the nuclear KE term from the total Hamiltonian. In our case
${\mathcal{H}}_{el}$ describes 
two electronic states separated by an energy gap that linearly depends on $Q$, 
according to $2z(Q) = 2z_0-\sqrt{2\epsilon_{sp}}\omega Q$.\cite{painelli98}
By diagonalizing the electronic $2\times2$ matrix  
one gets analytical expressions for the $Q$-dependent ground and excited
state energies, as shown in Fig.~1b. 
It is important to recognize that, even if the Hamiltonian
 in Eq.~(\ref{hamiltonian}) is defined in terms of harmonic reference states 
(Fig.~1a), as a consequence 
of e-ph coupling,  the ground and excited state  PES are anharmonic (Fig.~1b).
 The anharmonicity
of the potential prevents the analytical solution of the vibrational 
problem on either the ground or excited state; however numerically exact
vibrational states can be calculated in both manifolds.
In particular,
the eigenstates of the harmonic oscillator with frequency $\omega$, 
centered at the relevant equilibrium position, are a good basis
for the vibrational problem on either PES. The corresponding
 vibrational Hamiltonian
is the sum of a KE term, whose matrix elements are trivial in the
adopted basis, plus a PE term, whose matrix elements are calculated
via numerical integration. 
Of course the vibrational matrix is diagonalized on a  basis 
truncated to a   large enough number of phonon states as to 
get convergence. Results presented in this paper have typically been
obtained with 20 phonon states. Once BO eigenstates are 
obtained, the (transition) dipole moments entering SOS expressions
can be calculated via  numerical integration.

Static susceptibilities  calculated within BO approximation are
indistinguishable (in the scale of Fig.~2) from the exact ones,
as long as   $\omega\leq 0.2$.
Fig.~3 compares non-adiabatic (continuous lines) and BO (dashed lines) 
static susceptibilities  for $\omega =0.5$, where deviations appear. 
Of course BO approximation becomes worst with increasing $\omega$ and 
is totally untenable for $\omega\geq 1$. 
The adopted value of the small polaron binding energy, 
$\epsilon_{sp}\sim\sqrt{2}t$ applies to both push-pull 
chromophores\cite{kim,terencpl,jpc,goddard} and 
to CT complexes and  salts.\cite{bibbia,pecile}
In push-pull chromophores the typical value of $\sqrt{2}t\sim 1$~eV is
much larger than typical vibrational frequencies 
($\omega\sim 1000$~cm$^{-1}$)\cite{kim,terencpl,jpc,goddard}
and the dimensionless $\omega\sim 0.1 - 0.2$ is safely in the BO regime. 
In CT salts, instead, $\sqrt{2}t$ is much smaller 
($\sim 0.2$~eV)\cite{bibbia,pecile} and the corresponding 
dimensionless $\omega\sim 0.5 - 1$ suggests seizable  non-adiabatic effects
in NLO responses of these (narrow-band)  systems.

\section{SOS susceptibilities}

 The separation of electronic and vibrational degrees of freedom in 
BO approximation not only leads to simpler calculations,
but also allows for additional insight.
When applied to SOS expressions, BO quite immediately leads to a separation
of electronic and vibrational contributions. If $|Rv\rangle$ indicates a BO
vibronic state, i.e. the product between an electronic state $|R\rangle$ and a 
vibrational state $|v\rangle$ in the $R$ manifold, the SOS expressions for 
static susceptibilities read:\cite{orr}
\begin{eqnarray}
\alpha &=& 2\sum_{R,v} \frac{\langle 0G|\hat{\mu}|Rv\rangle 
\langle vR|\hat{\mu}|G0\rangle}{\omega_{Rv}} \nonumber \\
\beta &=& 6\sum_{R,v,S,u} {\frac{\langle 0G|\hat{\mu}|Rv\rangle 
\langle vR|\overline{\mu}|Su\rangle 
\langle uS|\hat{\mu}|G0\rangle}
{\omega_{Rv}\omega_{Su}}} \label{sos} \\
 & & \nonumber \\
\gamma  &=&  \nonumber 
\\ &24& \sum_{R,v,S,u,T,w}
{\frac{\langle 0G|\hat{\mu}|Rv\rangle 
\langle vR|\overline{\mu}|Su\rangle
\langle uS|\overline{\mu}|Tw\rangle 
\langle wT|\hat{\mu}|G0\rangle}
{\omega_{Rv}\omega_{Su}\omega_{Tw}}}  \nonumber \\
   &-& 24 \sum_{S,u,T,w}
{\frac{\langle 0G|\hat{\mu}|Su\rangle 
\langle uS|\hat{\mu}|G0\rangle
\langle 0G|\hat{\mu}|Tw\rangle 
\langle wT|\hat{\mu}|G0\rangle}
{\omega_{Su}^2\omega_{Tw}}},    \nonumber
\label{sos}
\end{eqnarray}
where the gs, $|G0\rangle$, is excluded from the summations,
$\omega_{Rv}$  is the frequency of the $|G0\rangle\rightarrow |Rv\rangle$ 
transition, and
$\overline{\mu}=\hat{\mu}-\langle 0G|\hat{\mu}|G0\rangle$.
In the two-state model, the summations on $R$, $S$ and $T$ only run on the two 
electronic states $|G\rangle$ and $|E\rangle$.
Then, according to the standard definition\cite{bishop91}, 
the electronic contribution
to susceptibilities is given by terms 
in Eq.~(\ref{sos}) with $R=S=T=E$, 
and the vibrational contribution is described by
terms where at least one of $R$, $S$ or $T$ states coincides with $|G\rangle$.
We underline that electronic contributions to BO-SOS 
susceptibilities do not  coincide with the responses of the bare 
electronic system (dot-dashed lines in Fig.~2). 
In fact summations in Eq.~(\ref{sos}) run on
 vibronic states and not on pure electronic states. In the reliability 
range of BO approximation, closure on vibrational states in the excited
state manifold works fine in reducing sums on intermediate vibronic states 
into  sums on pure electronic states, but the property is evaluated 
for the true vibronic gs ($|G0\rangle$) and not as a pure electronic property
(mediated over $|G\rangle$): the electronic contribution to BO-SOS 
susceptibilities partly include zero-point vibrational average (ZPVA)
corrections, as defined in Ref.\cite{bishop90}.

The CN approximation\cite{bishop98,bishop91,bishop90} introduces a different 
way of partitioning electronic and vibrational degrees of freedom.
In this approach the electronic contribution
to the $n$-th order susceptibility is calculated as  the $n$-th order
bare electron susceptibility, $\chi^{(n)}_0$.
Two more contributions are then added to get the total response: 
(a) ZPVA corrections, that  account for the difference between the electronic
susceptibility calculated at the bottom of the gs PES (i.e. at the equilibrium
geometry) and that relevant to the ground vibronic state (due to anharmonicity
the relevant geometry does not coincide with the equilibrium one).
These corrections are  usually 
calculated via an expansion of electronic susceptibilities on $Q$;\cite{bishop98} 
(b) vibrational contributions, that are calculated in terms of summations
running on  the vibrational states in the gs manifold only, as detailed 
in Ref.\cite{bishop90}.

BO-SOS and CN-SOS are based on the same adiabatic approximation and lead
to basically identical results at least as long as the adiabatic approximation 
itself is reliable (approximately $\omega < 0.5$). 
However, the partitioning of the susceptibilities
is different and
there is not a one-to-one correspondence between terms appearing in the two
schemes: CN-SOS
kills a few terms appearing in BO-SOS, and, at the same time, some of the
terms, that are usually considered as corresponding,\cite{bishop91} 
do not have exactly the  same meaning in the two approaches.
As a matter of fact, differences between CN and BO results amount to
$\sim 7\%$ and $\sim 15\%$ in the electronic
$\beta$ and $\gamma$, respectively, and $\sim 2\%$ and $\sim 3\%$
in the vibrational contributions to  $\beta$ and $\gamma$
(these estimates refer
to the parameters $\epsilon_{sp}=1$, $\omega=0.2$; for $\omega=0.1$ deviations
are even smaller).
These small differences  compensate each other, leading to negligible 
($\le$ 1\%) differences in the total (electronic+vibrational) responses.

Our numerical solution of the adiabatic problem for the two-state model
allowed us to test BO and CN approximations without introducing additional
approximations. However getting exact vibrational eigenstates 
is hardly possible for complex (molecular) structures. For this reason
CN-SOS calculations are often implemented by invoking the harmonic approximation 
for the vibrational problem.\cite{bishop98,bishop91,champagne94,bishopjcp}
As already stressed in the Introduction,
the harmonic approximation   fails in 
systems where vibrations are coupled to electrons with large nonlinear responses.
The relation between anharmonicity and  nonlinearity is very well apparent
in the adopted model, where both $Q$ and $F$ variables are coupled to the same
electronic operator, $\hat{\rho}$ (cf Eq.~(\ref{hamiltonian})).
Successive $Q$-derivatives of the ground
and excited state PE (${\mathcal E}_G$ and ${\mathcal E}_E$, respectively)
are therefore directly related to  
$F$-derivatives, i.e. to the bare electronic 
susceptibilities, $\chi_0^{(n)}$, as follows:
\begin{equation}
\frac{\partial^n{\mathcal{E}}_{G/E}}{\partial Q^n} =
\omega^2\delta_{2n}\mp
(2\epsilon_{sp})^{n/2}\omega^n\frac{\chi_0^{(n-1)}}{\mu_0^n},
\label{deriv1} 
\end{equation}
where $\delta_{2n}$ is the Kronecker-$\delta$, equal to 1 if $n=2$, zero
otherwise. 
From this equation it turns out that  systems with large 
{\it hyper}polarizabilities, i.e. with large  $\chi_0^{(n)}$, for $n\ge 2$,
are characterized by largely anharmonic  ground
and excited state PES. 
As it has been  discussed in Ref.\cite{delfreo,understanding}, 
the anharmonicity of vibrations
coupled to delocalized electrons is hardly detected in coherent
spectral measurements (e.g. electronic and vibrational absorption and/or 
Raman spectra), where experimental data can be satisfactorily reproduced in 
terms of parabolic PES with effective curvatures, but it shows up 
with large effects in incoherent spectral measurements (e.g. steady-state
and/or time-resolved emission).\cite{terencpl}
Large anharmonic corrections are also expected
in static nonlinear optical responses,\cite{delfreo,understanding} 
as it is confirmed by extensive
{\it ab initio} results on  conjugated materials.\cite{champagne00,luis}

The dotted lines in Fig.~2 show the ($\omega$-independent) susceptibilities
calculated in the best harmonic approximation (BHA)\cite{delfreo}.
In this approach the ground and excited state PES are approximated by
the two  parabolas that best fit the two anharmonic PES at the gs 
equilibrium position (i.e. at the minimum of the gs PES): 
the relevant curvatures are then defined by Eq.~(\ref{deriv1}) in terms
of the bare electronic linear polarizability evaluated at the equilibrium geometry.
Moreover, all electronic properties appearing in CN-SOS 
expressions are truncated to the linear term in $Q$.\cite{agren} 
We defer a detailed comparison of BHA and exact curves
to the next section, here we only underline that, as far as $\alpha$ is 
concerned, BHA and PE estimate exactly coincide, confirming that
in the $\omega \rightarrow 0$ regime anharmonic corrections
to the linear polarizability tend to vanish\cite{delfreo,agren}.
In the same limit,  nonlinear susceptibilities ($\beta$ and $\gamma$)
are largely amplified
by the anharmonicity of the gs PES, as demonstrated in panels (a) and (b)
of Fig.~2, by the large deviations 
of BHA results (dotted lines) from PE curves (dashed lines).

\section{Finite-field susceptibilities: 
potential and kinetic energy contributions}

In SOS approaches, susceptibilities are naturally partitioned  into 
electronic and vibrational contributions. Within FF approach, on the opposite,
PE contributions to the susceptibilities are naturally separated 
from contributions  due to the  nuclear KE. As already discussed in 
the Introduction, the lowest eigenstate of the electronic Hamiltonian 
as defined in the adiabatic approximation, is the
PE for the motion of nuclei (in the gs manifold, of course). 
The total gs energy is obtained by summing the nuclear KE to the PE.
Then, as long as BO applies, susceptibilities, i.e. the successive derivatives
of the gs energy with respect to an applied electric field, can be
calculated as sums of PE and KE $F$-derivatives. 
As a matter of fact, one can further separate the PE
contribution to susceptibilities into an electronic and a vibrational
part by simply comparing derivatives calculated at fixed nuclear
geometry with those taken by allowing
nuclei to relax following the application of the electric 
field.\cite{painelli98} This separation is however fairly artificial and 
does not add so much to our understanding.

The nuclear KE vanishes in the $\omega = 0$ limit, and the
($\omega$-independent)  PE susceptibilities, reported as dashed lines 
in Fig.~2, represent the zero-frequency limit of the exact susceptibilities.
Analytical expressions for the  PE-susceptibilities of the Holstein DA dimer
were already reported in Ref. \cite{painelli98}.
More generally, the calculation of PE-susceptibilities is easily implemented  
in quantum chemistry calculations,  in fact it only requires the gs energy 
calculated at the 
relaxed geometry for different values of an externally applied field.
Of course, geometry optimization is a crucial step in FF approach to
hyperpolarizabilities and it is extremely important to fully relax the geometry
for the chosen molecule in order to avoid  spurious `strain' or
`instability' contributions to hyperpolarizabilities.\cite{ingamells}

The calculation of KE contributions is more difficult,
since the $F$-dependence of the lowest vibrational state in the anharmonic
gs PES is needed. In the adopted model, however, the calculation
is feasible and, as long as BO applies, leads to basically exact results.
Specifically, much as it occurs for BO-SOS and CN-SOS, for $\omega\le 0.2$
exact (continuous) curves in Fig.~2 are indistinguishable from BO-FF results.
The nuclear KE contributes to susceptibilities in two different
ways. First of all, due to anharmonicity, the molecular geometry 
in the vibronic gs is different from the equilibrium geometry (corresponding
to the minimum of the gs PES). The PE susceptibilities have, in principle,
to be corrected to account for this effect. The correction is 
however very small, and we found it always negligible in the 
investigated parameter range. 
The second contribution stems from the $F$-dependence of the nuclear
KE itself: it is this contribution that indeed accounts 
for the deviations of the exact curves form the (dashed) PE curve (at least
in the BO regime, $\omega \le 0.2$, where non-adiabatic corrections are
negligible).
KE contributions  are of course very small for low $\omega$, 
but they  increase with  increasing 
$\omega$, leading to an overall decrease of the
vibrational amplification of the static NLO responses. 
This is by no means accidental: in the antiadiabatic limit 
($\omega\gg \sqrt{2}t$) phonons cannot contribute to static susceptibilities 
and,  with increasing $\omega$, KE contributions progressively
increase to counterbalance  PE contributions.  

KE contributions to susceptibilities exactly vanish in the harmonic approximation.
In fact, the equilibrium position in any harmonic vibrational state 
coincides with the bottom of the PES; moreover the nuclear KE is proportional 
to the harmonic frequency, i.e. to the curvature of the PES. For a parabolic
PES, this quantity is obviously independent of $Q$, and hence of $F$
(cf Eq.~(\ref{deriv1})).
We are then in the position to prove 
that the harmonic approximation is totally uncontrolled for the
calculation of {\it hyper}polarizabilities  in any frequency regime.
At low $\omega$, where PE contributions dominate, the harmonic approximation 
cannot account for anharmonic PES: this has no effect on the linear response,
and the BHA estimate of $\alpha$ coincides with the exact curve in the 
$\omega =0$ limit.\cite{delfreo} 
Anharmonic contributions to PE-{\it hyper}polarizabilities are instead
large, as demonstrated by the large deviations of BHA curves from the 
$\omega =0$ limiting curves.
At the same time, not accounting for KE corrections, harmonic results do not  
represent a reliable approximation in the large $\omega$ regime either. 
It is possible that in a given parameter range BHA gives good
estimates of a specific response, but this is due to an accidental
compensation of two large errors. The harmonic approximation is unreliable in
the calculation of {\it hyper}polarizabilities, and, 
as it turns out clearly from Fig.~2, it is not
possible to devise any parameter range where harmonic results represent a good 
approximation for all static susceptibilities.

\section{Conclusions}

The Mulliken $DA$ dimer model, extended to account for the Holstein coupling
to vibrations (or equivalently to other slow degrees of freedom,
including the orientational solvation coordinate), contains the essential 
physics to understand the subtle interplay between electrons, vibrations and 
external perturbations (namely an electric field) governing NLO responses
of molecular materials.\cite{painelli98}  This simple toy-model is then very useful to test
the reliability of several approximation schemes usually introduced
in more refined quantum chemical description of these materials. 
It is important to recognize that some very well-known and very widely
applied approximations can fail when applied to model the properties
of materials with strongly nonlinear behavior.
Perturbation theory, and particularly linear perturbation theory, is 
clearly inadequate in this context, and we already proved its inadequacy
in reproducing spectral properties of push-pull chromophores\cite{understanding,terencpl,jpc} as well as their
static susceptibilities.\cite{painelli98} More subtle effects have been demonstrated in 
two-photon absorption (TPA) spectra, where the standard Condon approximation
badly fails even when it properly reproduces one-photon absorption
spectra.\cite{tpa} Even more interestingly, an important vibrational channel 
is found to contribute to TPA spectra, in 
addition  to standard electronic channels, leading to large effects in
observed spectra.\cite{tpa}  The harmonic approximations works fine in TPA
spectra and more generally in vertical (coherent) processes,
but it is inadequate to describe incoherent processes.\cite{understanding}

In this paper we have addressed 
the approximations usually introduced in the calculation
of static NLO responses. The adiabatic approximation 
is fairly safe, at least as long as phonon frequencies are not too high
(approximately $\omega <  0.5\, t$, cf Fig.~3). It can be applied in different
ways, BO or CN, in SOS approaches leading to a slightly different partition
of electronic and vibrational contributions to susceptibilities,
but to the same overall result. In FF it leads to a partitioning 
of potential and kinetic energy contributions to susceptibilities, 
but otherwise leads to the same result as in SOS. Closure over
vibrational states is also a very good approximation, as long as the
adiabatic approximation applies.

Potential energy contributions are clearly dominant over kinetic energy
contributions in the low-vibrational frequency regime. 
The anharmonicity of the gs PES
therefore gives very large corrections to {\it hyper}polarizabilities: the harmonic 
approximation is untenable in this regime. With increasing phonon frequency, 
kinetic energy contributions globally reduce vibrational corrections 
until, in the antiadiabatic limit ($\omega \rightarrow \infty$) only 
electrons respond to static fields to regain purely electronic 
susceptibilities. Kinetic energy corrections vanish in the harmonic 
approximation and this approximation fails in the high frequency regime too. 

Both FF and SOS approaches can be easily implemented in quantum chemistry
calculations: the former has the advantage of only requiring knowledge about
the lowest eigenstate, whereas all eigenstates are required in SOS. 
However only $F=0$ eigenstates are needed in SOS, whereas the $F$-dependence
of the gs energy (or dipole moment) is required in FF. 
Anharmonicity is important and must 
be accounted for: in the low-vibrational frequency limit, where
 kinetic energy contributions are negligible, FF calculation is very convenient
and fully accounts for anharmonicity. 
In the intermediate frequency regime the PE-FF calculation overestimates
the vibrational amplification of NLO responses: either KE corrections
have to be introduced in the FF scheme, or one has to make resort to 
SOS calculations, but in any case anharmonicity has to be accounted in
order to get reliable estimates of {\it hyper}polarizabilities.

\begin{centerline}
	{ACKNOWLEDGEMENT}
\end{centerline}
Work supported by the Italian National Research Council (CNR)
 within its ``Progetto Finalizzato
 Materiali Speciali per Tecnologie Avanzate II'' and by the Ministry
of University of Scientific and Technological Research (MURST).

\vfill
\pagebreak

\onecolumn

\begin{figure}
\begin{center}
\mbox{\epsfig{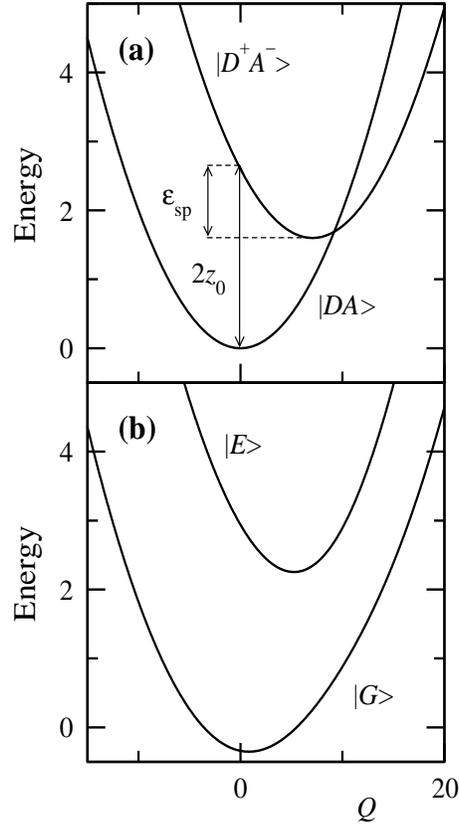}}
\end{center}
 \caption{Potential energy surfaces 
 for (a) the basis states 
($\sqrt{2}t =0$) and (b) the exact eigenstates ($\sqrt{2}t =1$)
of the Hamiltonian in Eq. (1). All curves are calculated for 
$z_0=1.3$ and $\epsilon_{sp}=1$.}
\end{figure}

\pagebreak
\twocolumn

\begin{figure}
\begin{center}
\mbox{\epsfig{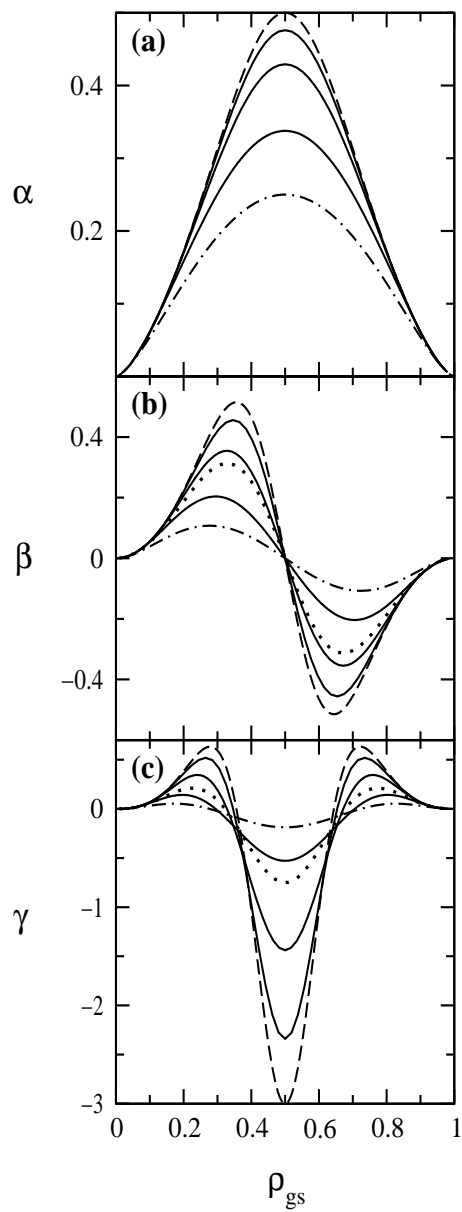}}
\end{center}
 \caption{Static (hyper)polarizabilities as a function of the ground state
ionicity $\rho_{gs}$, 
for $\epsilon_{sp} =1$ and different $\omega$ values ($\sqrt{2}t$ units).
Dashed lines: potential energy contribution to susceptibilities, 
corresponding to $\omega = 0$ limit (see text);
dot-dashed lines: bare electronic susceptibilities, corresponding
to the $\omega\rightarrow \infty$ limit (see text);
continuous lines report exact susceptibilities calculated for 
$\omega=0.05$, 0.2, 1.0, 
smoothly evolving from the $\omega= 0$ to the
$\omega \rightarrow  \infty$ limits;
dotted lines report susceptibilities calculated in the best harmonic approximation
(see text).
In panel (a), dotted and dashed lines are exactly superimposed. 
For the calculation of $\alpha$, $\beta$ and $\gamma$, 
dipole moments have been expressed in $\mu_0$ units.}
\end{figure}

\begin{figure}
\begin{center}
\mbox{\epsfig{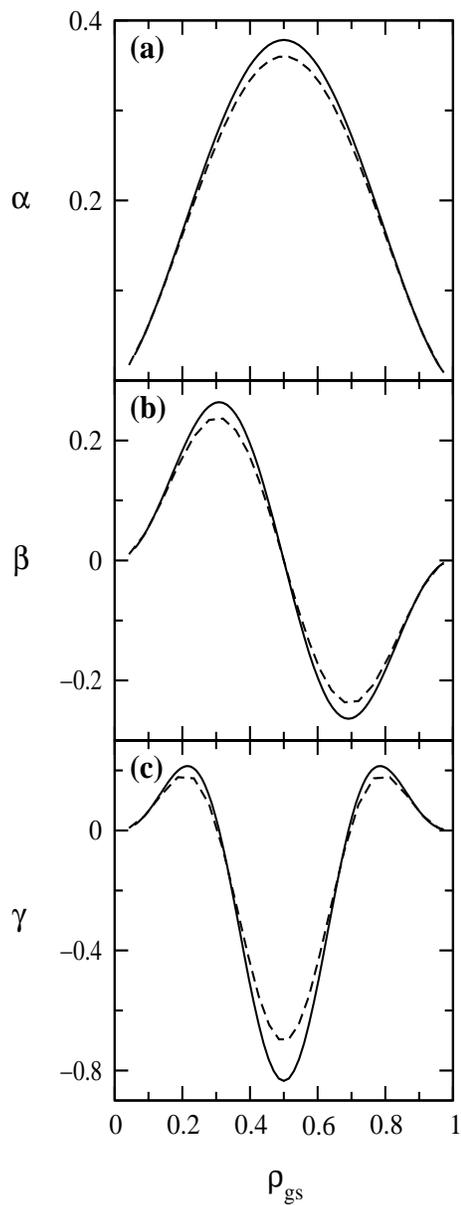}}
\end{center}
\caption{Static (hyper)polarizabilities as a function of the ground state
ionicity $\rho_{gs}$, for $\epsilon_{sp} =1$ and $\omega=0.5$ ($\sqrt{2}t$ units).
Continuous lines show non-adiabatic results; dashed lines show
BO results.}
\end{figure}

 \vfill
 \eject
 
 \end{document}